\documentclass[notitlepage,a4paper,aps,prd,onecolumn,superscriptaddress,nofootinbib,groupedaddress]{revtex4}
\usepackage{enumerate}
\usepackage{amsmath}
\usepackage{amsfonts}
\usepackage{amssymb}
\usepackage[utf8]{inputenc}
\usepackage[T1]{fontenc}
\usepackage{mathtools}
\usepackage{wasysym}
\usepackage{accents}
\usepackage[svgnames]{xcolor}
\usepackage[colorlinks=true, pdfstartview=FitV, linkcolor=blue, citecolor=red, urlcolor=magenta]{hyperref}
\usepackage{url}
\newcommand{\be}{\begin{equation}}
\newcommand{\ee}{\end{equation}}
\newcommand{\bea}{\begin{eqnarray}}
\newcommand{\eea}{\end{eqnarray}}

\usepackage{bibentry}
\usepackage{natbib}

\begin{document}

\title{The extended phase space thermodynamics of Planck-scale-corrected Reissner-Nordstr\"{o}m-anti-de Sitter black hole}

\newcommand{\JPess}{
\affiliation{Physics Department, Federal University of Para\'iba, Caixa Postal 5008, 58059-900, Jo\~ao Pessoa, PB, Brazil}
}

\newcommand{\Areia}{
\affiliation{Department of Chemistry and Physics, Federal University of Para\'iba, Rodovia BR 079 - Km 12, 58397-000 Areia-PB,  Brazil}
}

\newcommand{\ufrj}{
\affiliation{Instituto de F\'isica, Universidade Federal do Rio de Janeiro, 21.941-972 - Rio de Janeiro, RJ, Brazil}
}

\newcommand{\RIAMM}{
\affiliation{Research Institute for Astronomy and Astrophysics of Maragha (RIAAM), University of Maragheh, P.O. Box 55136-553, Maragheh, Iran}
}

\newcommand{\Lavras}{
\affiliation{Physics Department, Federal University of Lavras, Caixa Postal 3037, 37200-000 Lavras-MG, Brazil}
}

\author{Iarley P. Lobo}
\email{lobofisica@gmail.com}
\email{iarley\_lobo@fisica.ufpb.br}
\Areia
\Lavras
\author{Luis C. N. Santos}
\email{luis.santos@ufsc.br}
\JPess
\author{V. B. Bezerra}
\email{valdir@fisica.ufpb.br}
\JPess
\author{J. P. Morais Gra\c ca}
\email{jpmorais@gmail.com}
\ufrj
\author{H. Moradpour}
\email{hn.moradpour@maragheh.ac.ir}
\RIAMM

\begin{abstract}
We analyze the effect of Planck-scale modified radiation equation
of state on the Reissner-Nodstr\"{o}m-anti-de
Sitter black hole inspired by Kiselev's ansatz. Deformed
thermodynamic quantities are found, phase transitions and black
holes as heat engines are described for the Carnot and square
cycles. Non-trivial differences between linear and quadratic
Planck-scale corrections are discussed in detail.
\end{abstract}


\maketitle


\section{Introduction}
A consistent quantization of gravitational degrees of freedom has
been one of the main quests of theoretical physics in the recent decades. Many proposals based on very
different approaches to this problem have emerged, where each of
them is responsible for filling an aspect of this general problem.
Among some of the most prominent programs we highlight, for
instance, loop quantum gravity (LQG) \cite{Rovelli:1997yv}, causal
dynamical triangulation \cite{Ambjorn:2010rx}, causal sets
approach \cite{Henson:2006kf}, string theory \cite{Mukhi:2011zz}.
The maturity of the quantum gravity problem and technological
advances have led the community to propose phenomenological
approaches that can capture some of the main properties of
the proposed solutions of this problem in a bottom-up way, where deformations of known concepts and equations of
general relativity and quantum mechanics are considered, governed by parameters that can become important when quantum
gravitational effects appear (we refer the reader
to the review \cite{AmelinoCamelia:2008qg}).
\par
Amongst these phenomenological proposals, some of the main sources of constraints are provided by deformations or violations of the Lorentz transformations, induced by the quantum gravity energy scale (supposedly of the order of the Planck energy) \cite{Ackermann:2009aa,Acciari:2020kpi,Albert:2019nnn}. In summary, the interaction of particles with quantum gravitational degrees of freedom could be described kinematically by a modified Hamiltonian or dispersion relation, such that trajectories would be modified and symmetries could be deformed. Besides that, a fruitful way of realizing the kinematics of the particles endowing corrections in intermediary regime between the full quantum gravity theory and special/general relativity consists in describing their motion as if they propagate in a spacetime that depends on properties of individual particles themselves, like the ratio between their energy/momentum and Planck energy \cite{Magueijo:2002xx,Girelli:2006fw,Amelino-Camelia:2014rga,Lobo:2016xzq,Lobo:2016lxm,Lobo:2020qoa,Barcaroli:2015xda,Weinfurtner:2008if,Assaniousssi:2014ota,Ling:2005bp,Ling:2005bq,Li:2008gs}, or properties of fluids of particles, like the ratio between its energy density and Planck energy \cite{Olmo:2011sw} or the fluid's temperature and Planck temperature \cite{Lobo:2019jdz}.
\par
These phenomenologically-inspired modifications have been demonstrated to be a useful tool able to scrutinize the Planckian regime without the need of passing through specific details of each quantum gravity proposal, i.e., this approach allows one to derive known Planck-scale corrections to some expressions, for instance, the logarithmic correction of Hawking-Bekenstein black hole entropy formula \cite{AmelinoCamelia:2004xx,Amelino:2006}, to regularize the behavior of a black hole during its final stages of existence \cite{Adler:2001vs}, or to describe cosmological bounces derived from loop quantum cosmology \cite{Gorji:2016laj}.
\par
When applying such procedure to the photon gas problem, it has been long known that one detects non-trivial modifications of the equation of state parameter \cite{Amelino:2006} proportional to the ratio between the temperature of the fluid and Planck temperature. On the other hand, it has been found by Kiselev a solution of the general relativity field equations \cite{Kiselev:2002dx} that is able to modify the static black hole metric in a way that depends on an averaged equation of state parameter of the fluid that surrounds the black hole. In particular, the Reissner-Nordstr\"{o}m solution has been shown to correspond to the special case in which the average equation of state parameter corresponds to that of a photon gas. Planck-scale corrections of the Reissner-Nordstr\"{o}m metric induced by the above mentioned modified thermodynamics have been recently analyzed by some of the authors in \cite{Lobo:2019jdz}. Besides that, the impact of the trans-Planckian regime and the novel notion of thermal dimension \cite{Amelino-Camelia:2016sru,Husain:2013zda,Nozari:2015iba} on the evaporation of charged black holes have been studied in \cite{Lobo:2020oqb}.
\par
On the other hand, the Reissner-Nordstr\"{o}m black hole has proven to be an important tool for research in thermodynamical aspects of gravity, when considered along with a negative cosmological constant. In fact, the extended phase space thermodynamics of the Reissner-Nordstr\"{o}m-anti-de Sitter black hole has been analyzed in recent years from the seminal papers \cite{Kastor:2009wy,Kubiznak:2012wp}, and from then it has been explored in a large amount of scenarios \cite{Brihaye:2018nta,Ovgun:2017bgx,Ma:2017pap,Upadhyay:2017fiw,Fernando:2016sps,Hendi:2015kza,Dehghani:2014caa,Mo:2014mba,Hendi:2016njy,Liu:2016uyd,Majhi:2016txt,kiselevgb}, in which the main issue under analysis are phase transitions and heat engines \cite{Johnson:2014yja} induced by the presence of the electric charge or some other terms due to different matter contents surrounding the black hole, like clouds of strings, quintessence, etc, or alternative theories of gravity \cite{Balart:2019uok,Zhang:2018hms,Fernando:2018fpq,EslamPanah:2018ums,Mo:2018hav,Rosso:2018acz,Zhang:2018vqs,Chakraborty:2017weq,Wei:2017vqs,Hendi:2017bys,Xu:2017ahm,Liu:2017baz,Hennigar:2017apu,Chakraborty:2016ssb,Zhang:2016wek,Bhamidipati:2016gel,Johnson:2015fva,Johnson:2015ekr,Setare:2015yra,Lobo:2019put} (see \cite{Kubiznak:2016qmn} for a comprehensive review on the subject).
\par
Therefore, a natural question is imposed: what is the effect of quantum corrections induced by departures of the classical equation of state parameter of the radiation fluid $\omega=1/3$ in the scenario of black holes in the extended phase space that includes a negative cosmological constant, i.e., how are affected the thermodynamic quantities, phase transitions and efficiency of this system operating as a heat engine? Answering these questions is the main goal of this paper.
\par
In this paper, we explore the non-trivial effect of pressure induced by a negative cosmological constant on the thermodynamics of the Planck-scale modified Reissner-Nordstr\"{o}m-anti-de Sitter spacetime. In section \ref{sec:kiselev}, we revisit Kiselev's solution of a black hole surrounded by matter content with an average equation of state parameter. In section \ref{sec:state}, we calculate modifications in the equation of state parameter induced by Planck-scale modified dispersion relation. In section, \ref{sec:therm}, we derive the thermodynamic quantities in the extended phase space assuming the results of the previous sections and a negative cosmological constant. In section \ref{sec:crit}, we describe the impact of the quantum gravity parameter on phase transitions. In section \ref{sec:engine}, we calculate the effect of the Planck scale parameter on the efficiency of the black hole system when working as a heat engine. We draw our final remarks in section \ref{sec:conc}.
\par
We assume $c=\hbar=k_B=1$, for simplicity, and metric signature $(+---)$.

\section{Kiselev's solution}\label{sec:kiselev}

The solution of general relativity's field equations describing a static, spherically symmetric black hole surrounded by an average quintessence fluid was found in the seminal paper \cite{Kiselev:2002dx} by Kiselev, in which the metric and stress energy-tensor of the surrounding matter is given by
\bea
ds^2=A(r)dt^2-\frac{dr^2}{A(r)}-r^2d\Omega^2\, ,\\
T^t{}_t=T^r{}_r=-\rho(r)\, ,\\
T^{\theta}{}_{\theta}=T^{\phi}{}_{\phi}=\frac{1}{2}(3\omega+1)\rho(r)\, ,
\eea
where $A(r)$ is a function of the radial coordinate $r$, $\rho(r)$ is the energy density distribution and $\omega$ is an equation of state parameter that relates the average pressure and energy density. This procedure allows us to effectively capture the physical properties of the surrounding fluid and will be of great usefulness for our purposes.

From Einstein's field equations,
\be
G_{\mu\nu}=\kappa T_{\mu\nu}\, ,
\ee
we are able to find the shape of the energy-density profile $\rho(r)$ and the metric function $A(r)$:
\bea
\rho(r)=\frac{c}{\kappa}\frac{3w}{r^{3(w+1)}}\, ,\\
A(r)=1-\frac{2GM}{r}+\frac{c}{r^{3w+1}}\, .
\eea

From equations above, one can recognize a generalization of the Schwarzschild metric, whose new contribution is null when the surrounding energy density is absent ($c=0$) and depends on the average equation of state parameter $\omega$, that labels different fluids that surround the black hole. This solution obeys an additivity condition such that different fluids can, in principle, contribute to the stress-energy tensor, which reflects in contributions to the metric as $\sum_n(r_n /r)^{3\omega_n+1}$. From this approach it is possible to reconstruct the Reissner-Nordstr\"{o}m-anti-de Sitter spacetime from contributions of a negative cosmological constant with equation of state parameter $\omega_1=-1$, and a radiation fluid, which in principle presents an equation of state parameter $\omega_2=1/3$. This gives
\be
A(r)=1-\frac{2GM}{r}-\frac{\Lambda}{3}r^2+\left(\frac{Q}{r}\right)^2\, .\\
\ee

As we will see in the following sections, the presence of corrections in the radiation equation of state parameter due to Planck scale effects, $\omega\approx 1/3+{\cal O}(T_P)$, (where $T_P$ is the Planck temperature) will allow us to derive non-trivial corrections to the extended phase space thermodynamics.


\section{Planck-scale-deformed equation of state parameter}\label{sec:state}

In this section, we will revisit the problem of the photon gas by calculating its internal energy and pressure from the occupation number of states with given momenta and the partition function of this bosonic system. As we shall see, Planck-scale corrections shall modify the relation between momenta and energy, which will lead to a deformation of the equation of state parameter.

In fact, we can describe a special class of isotropic modified
dispersion relations of particles with mass $m$ by relying on
deformation functions $f(E/E_{QG})$ and $g(E/E_{QG})$ as follows
\begin{equation}\label{mdr1}
m^2=E^2f^2(E/E_{QG})-p^2g^2(E/E_{QG})\, ,
\end{equation}
where $p$ is the particle's momentum, $E$ is its energy and $E_{QG}$ is the quantum gravity energy scale, typically assumed to be of the order of the Planck energy $E_P\approx 1.22\times 10^{19}\, \text{GeV}$.

These Planck-scale deformations are known to
induce deformations in the behavior of various
gases with several applications from black hole physics to
cosmology \cite{Santos:2015sva,Nozari:2006yia,Alexander:2001ck,Alexander:2001dr,Lobo:2020oqb,Moradpour:2021ymp},
where the nontrivial rule for counting states within a given
energy/momentum interval leads to interesting phenomena, like
inflation without an inflaton field and dimensional reduction due
to quantum gravity at the trans-Planckian regime.
\par
The number of states with momentum between $p$ and $p+dp$ contained in a volume $V$ is given by \cite{huang}
\begin{equation}
N(p)dp=\frac{V}{(2\pi)^3}4\pi p^2 dp\, ,
\end{equation}
and the relation between the momentum and the energy is given by the deformed expression (\ref{mdr1}) when $m=0$. Assuming the degeneracy due to the two polarizations of the photons, from Eq.(\ref{mdr1}) a straightforward calculation implies that \cite{Lobo:2019jdz,Santos:2015sva}
\par
\begin{equation}
2N(p)dp\doteq 2\tilde{N}(E)dE=2\frac{V}{\pi^2}\left(\frac{f}{g}\right)^3\left(1+E\frac{f'}{f}-E\frac{g'}{g}\right)E^2dE\, ,
\end{equation}
where prime ($'$) denotes differentiation with respect to the energy.
\par
As usual, the equation of state parameter can be found from the average of the energy density and the pressure, for which the Bose-Einstein statistical distribution remains unmodified \cite{Alexander:2001ck}:
\begin{equation}\label{omega1}
\omega(T)=\frac{P}{\rho}=-T\frac{\int \ln[1-e^{-E/T}]\tilde{N}(E)dE}{\int\frac{E}{\exp [E/T]-1}\tilde{N}(E)dE}\, ,
\end{equation}
where $T$ is the fluid's temperature. For an undeformed dispersion relation, each of these integrals are proportional to the $4$-th power of the temperature, which furnish an equation of state parameter given by the constant value of $1/3$, which corresponds to the usual value associated to a radiation fluid.
\par
We assume a MDR of the form (\ref{mdr1}), for instance, one that has been considered in several phenomenological scenarios of quantum gravity (as a way how to capture a upper limit on the energy of fundamental particles when $\xi <0$ \cite{Ling:2005bp,Ling:2005bq,Li:2008gs}):
\begin{equation}\label{mdr}
f=1 \ \ \ , \ \ \ g=\sqrt{1+\xi \left(\frac{E}{E_P}\right)^n}\, ,
\end{equation}
where $\xi$ is a dimensionless real parameter that labels different phenomenological proposals, and $n$ is a natural number that describes the leading order correction proportional to the ratio $(E/E_P)^n$. In this paper, we are considering the first and second order cases, i.e., $n=1$ and $n=2$, since these are the orders of magnitude that have been mostly prominent for observational constraints \cite{Acciari:2020kpi,Amelino-Camelia:2016ohi}, for simplicity. As we shall see, this difference will be enough to produce significantly distinctive scenarios.
\par
By following the above procedure and considering only the first term in a Taylor perturbation around $E/E_P$ in Eq.(\ref{omega1}), we find a temperature-dependent equation of state parameter:
\be\label{w1}
\omega_n(T)\approx\frac{1}{3}+\alpha_n\left(\frac{T}{T_P}\right)^n\, ,
\ee
where $T_P$ is the Planck temperature, $\alpha_1=\xi\, 60\zeta(5)/\pi^4$ and $\alpha_2=\xi\, 40\pi^2/63$ are dimensionless parameters that we shall allow to vary in our analysis, and $\zeta(x)$ is Riemann zeta function, which are typically expected in first order perturbations of the dispersion relation \cite{Lobo:2019jdz}. Now we proceed to applications in the Reissner-Nordstr\"{o}m-anti-de Sitter spacetime.


\section{Extended phase space thermodynamics}\label{sec:therm}

The individual interaction of high energetic photons with a quantum spacetime could be effectively described by the kinematics induced by a modified dispersion relation, in such a way that when extended to a collection of particles like a fluid, manifests as a temperature-dependent equation of state parameter. Such temperature is the one assigned to this high energetic radiation fluid. Now, if we suppose that such high energetic particles surround a static, spherically symmetric configuration, in a similar way to the one described in section \ref{sec:kiselev}, we shall expect Planck scale deformations of Kiselev's approach to the Reissner-Nordstr\"{o}m-anti-de Sitter spacetime.
\par
This approach allows to propose an effective way to describe a spacetime endowed with these properties that resembles a known approach called ``rainbow gravity'' \cite{Magueijo:2002xx}. In that case, the propagation of individual particles on a curved spacetime could be described by a metric that depends on the energy/momentum of the particles itself. More recent versions of this idea are realized by Finsler \cite{Girelli:2006fw,Amelino-Camelia:2014rga,Lobo:2016xzq,Lobo:2016lxm} and Hamilton geometries \cite{Barcaroli:2015xda}. Here, we propose that the spacetime probed by a fluid of massless particles could be described in a temperature-dependent way $g_{\mu\nu}(T/T_P)$, whose realization can be done by the Kiselev approach.
\par
In fact, we realize this idea by having a black hole spacetime of mass $M$ surrounded by a negative cosmological constant  $\Lambda$ and a radiation fluid defining a deformed Reissner-Nordstr\"{o}m-anti-de Sitter black hole of charge $Q$:
\be
A(r)=1-\frac{2GM}{r}-\frac{\Lambda}{3}r^2+\left(\frac{Q}{r}\right)^{3\omega(T)+1}\, ,
\ee
that depends on the ratio $T/T_P$ induced by the equation of state parameter (\ref{omega1}). The main difference of this approach in comparison to the previous ones consists in the fact that we are inducing Planck scale correction in geometry through the matter sector of the gravitational field equations, instead of accommodating the kinematics of particles in a certain non-Riemannian spacetime.
\par
In order to connect with black hole thermodynamics, we realize that the photons emitted via Hawking radiation present energies of the order of the black hole's temperature $E\sim T$ \cite{Ling:2005bp,Adler:2001vs}. For this reason, we analyze a stable scenario in which the temperature of the radiation that surrounds the black hole is of the order of the Hawking temperature as well. This is a common ansatz assumed in energy-dependent metric approaches \cite{Ling:2005bp} and also for quantum-gravity-induced generalized uncertainty principles \cite{Adler:2001vs}. This allows us to find the Hawking temperature as a function of the outer horizon radius $r_+$ (solution of $A(r_+)=0$):
\bea
T_H(r_+)=\frac{1}{4\pi}\frac{\partial}{\partial r}A(r_+)=\frac{1}{4\pi}\left[\frac{1}{r_+}+8\pi P r_+-\frac{3\omega(T)}{r_+}\left(\frac{Q}{r_+}\right)^{3\omega(T)+1}\right]\, ,\label{t1}
\eea
where we have, in fact, the temperature implicitly defined, since we are making $T=T_H$ and the pressure is identified from the cosmological constant, as usual in the extended phase space thermodynamics \cite{Kubiznak:2016qmn}, as
\be
P=-\frac{\Lambda}{8\pi}\, .
\ee

Concretely, from the dispersion relations used in the previous section, we use Eq.(\ref{w1}) in Eq.(\ref{t1}), replacing $\omega\rightarrow \omega_n$, and expanding in powers of the inverse of the Planck temperature, to find
\be\label{T}
T=T^{\circ}-\alpha_n\left(\frac{T^{\circ}}{T_P}\right)^n\frac{Q^2}{V^{\circ}}\left[1+\ln \left(\frac{Q}{r_+}\right)\right]\, ,
\ee
where $V^{\circ}=4\pi r_+^3/3$ is the Euclidean, undeformed volume and
\be\label{tcirc}
T^{\circ}=\frac{1}{4\pi}\left[\frac{1}{r_+}+8\pi P r_+-\frac{Q^2}{r_+^3}\right]
\ee
is the undeformed temperature of the Reissner-Nordstr\"{o}m-anti-de Sitter black hole.

This ansatz for the temperature modifies the thermodynamics of the black hole as we shall see. In fact, the mass $M$, that plays the role of enthalpy of this thermodynamic system, is defined by:
\be\label{mass1}
GM(r_+,P,Q)=\frac{r_+}{2}+\frac{Q^2}{2r_+}+\frac{4\pi}{3}Pr_+^3+\frac{3\alpha_nQ^2}{2r_+}\left(\frac{T^{\circ}(r_+,Q,P)}{T_P}\right)^n\ln\left(\frac{Q}{r_+}\right)\, ,
\ee
where $T^{\circ}=T^{\circ}(r_+,Q,P)$ is assumed as a function of the horizon radius, charge and pressure according to (\ref{tcirc}).

We describe the variation of the black hole's mass due to variations of the horizon radius $r_+$, pressure $P$ and electric charge $Q$ as usual
\be
dM=\frac{\partial M}{\partial r_+}\Bigg|_{P,Q}\delta r_+ + \frac{\partial M}{\partial Q}\Bigg|_{r_+,P}\delta Q + \frac{\partial M}{\partial P}\Bigg|_{r_+,Q}\delta P.
\ee

We also define the entropy, as usual, for static black holes from variation of the enthalpy with respect to the horizon radius as follows:
\be\label{ds}
dS\doteq \frac{1}{T}\frac{\partial M}{\partial r_+}\Bigg|_{P,Q}d r_+\, .
\ee
\par
This leads to the following expression for the \textit{linear correction} ($n=1$) after integration:

\begin{align}
S_1=\frac{\pi r_+^2}{G}+\frac{3\alpha _1}{2\sqrt{2}G\, T_P}\frac{Q}{r_+}\left\{3 \sqrt{2} Q \left[\ln \left(\frac{Q}{r_+}\right)-1\right]-2 r_+ \sqrt{a+1} \ln \left(\frac{Q}{r_+}\right) \tanh ^{-1}\left(\frac{4 r_+\sqrt{\pi P}}{\sqrt{a-1}}\right)\right.\\
+2 r_+ \sqrt{a-1} \ln \left(\frac{Q}{r_+}\right) \tan ^{-1}\left(\frac{4 r_+\sqrt{\pi P}}{\sqrt{a+1}}\right) -2r_+\sqrt{a+1}\, \chi_2\left(\frac{4 r_+\sqrt{\pi P}}{\sqrt{a-1}}\right)\left.+2r_+\sqrt{a-1}\text{Ti}_2\left(\frac{4 r_+\sqrt{\pi P}}{\sqrt{a+1}}\right)\right\}\, ,\nonumber
\end{align}

where $\alpha_1=\xi 60\zeta(5)/\pi^4$, $a=\sqrt{1+32\pi PQ^2}>1$, and the terms $\chi_{\nu}(z)=\frac{1}{2}[\text{Li}_{\nu}(z)-\text{Li}_{\nu}(-z)]$ and $\text{Ti}_2(z)=\frac{1}{2i}[\text{Li}_{2}(iz)-\text{Li}_{2}(-iz)]$ are called Legendre's chi function and inverse tangent integral function, respectively, and are defined in terms of the polylogarithm function $\text{Li}_{\nu}(z)=\sum_{k=1}^{\infty}z^k/k^{\nu}$. This expression reduces to the one found in \cite{Lobo:2019jdz} when $P\rightarrow 0$.
\par
This entropy $S_1$ presents the peculiar property of assuming complex values for a non-null pressure $P>0$. The concept of complex entropy has been recently studied in the literature \cite{c:entropy1,c:entropy2,c:entropy3} and it is suggested that its imaginary part describes the entropy transferred to the background, which would mean that such black hole is an unstable system \cite{c:entropy1}. This dissipation is driven by the pressure term given by the negative cosmological constant, i.e., when $P\rightarrow 0$, we recover a real valued entropy and stable system as considered in \cite{Lobo:2019jdz}. The conservation of energy described by the first law of thermodynamics needs to involve also the energy lost due to the pressure term. This information is already included in the nature of the entropy. As we shall see, this is a property that shall not be present when we assume quadratic terms as the leading order corrections.
\par
In fact, for $n=2$, we have the following entropy after integrating Eq.(\ref{ds}):
\bea
S_2=\frac{\pi r_+^2}{G}-\frac{3\alpha_2}{64\pi G\, T_P^2}\frac{Q^2}{r_+^4}\left\{4r_+^2-3Q^2+4\left[3Q^2-2r_+^2+16\pi P\, r_+^4\ln\left(\frac{Q}{r_+}\right)\right]\ln\left(\frac{Q}{r_+}\right)\right\}\, ,
\eea
where $\alpha_2=\xi 40\pi^2/63$. This is a real-valued quantity that reduces to the case studied in \cite{Lobo:2019jdz} when $P\rightarrow 0$. In these two cases, we see logarithmic corrections to the entropy, whose Planck-scale contributions vanish when $Q\rightarrow 0$.
\par
The electric potential for $n=1$ and $n=2$ reads:
\bea
\Phi_n\doteq\frac{\partial M}{\partial Q}\Bigg|_{r_+,P}=\frac{Q}{G r_+}+\frac{3\alpha_nQ}{2r_+}\frac{\left(T^{\circ}\right)^{n-1}}{G\, T_P^n}\left[T^{\circ}+\ln\left(\frac{Q}{r_+}\right)\left(2T^{\circ}-\frac{nQ^2}{2\pi r_+^3}\right)\right]\, ,
\eea
thus also presenting logarithmic corrections and, as usual, $T^{\circ}$ is a function of $r_+$, $Q$ and $P$, and $\alpha_n$ are given by the  quantities below Eq.(\ref{w1}).
\par
Besides that, a notion that is common in this so called black hole chemistry approach is the existence of a thermodynamic volume, which is read from the first law of thermodynamics and not necessarily is given by the Euclidean volume for cases beyond general relativity (for further discussions on this topic, we refer the reader to the review \cite{Kubiznak:2016qmn}). In our case, the thermodynamic volume also gains logarithmic correction as
\be
V\doteq\frac{\partial M}{\partial P}\Bigg|_{r_+,Q}=\frac{4\pi r_+^3}{3G}+\frac{3n\alpha_n Q^2}{G\, T_P^n}\left(T^{\circ}\right)^{n-1}\ln\left(\frac{Q}{r_+}\right)\, .
\ee
\section{Critical behavior}\label{sec:crit}
Endowed with these quantities, we can study how quantum corrections affect phase transitions due to the electric charge. As a first step, we need to establish an equation of state for the pressure, which can be straightforwardly calculated from Eqs.(\ref{T}) and (\ref{tcirc}):
\be\label{eq:state}
P=\frac{T}{2r_+}-\frac{1}{8\pi r_+^2}+\frac{Q^2}{8\pi r_+^4}+\frac{3\alpha_nQ^2}{8\pi r_+^4 }\left(\frac{T}{T_P}\right)^n\left[1+\ln \left(\frac{Q}{r_+}\right)\right]\, .
\ee
We find the critical quantities $(r_c,T_c,P_c)$ by solving the following system of equations
\be
\frac{\partial P}{\partial r_+}\Bigg|_{T,Q}=0=\frac{\partial^2 P}{\partial r_+^2}\Bigg|_{T,Q}\, ,
\ee
where $P_c\doteq P(r_+=r_c,T=T_c)$. This gives the following solution:
\begin{align}
T_c=\frac{\sqrt{6}}{18\pi Q}-\frac{\alpha_n}{A_n\, T_P^n}\left[5-\ln (36)\right],\\
r_c=\sqrt{6}Q+\frac{\alpha_n}{B_n\, T_P^n}\left[19-\ln (46656)\right],\\
P_c=\frac{1}{96\pi Q^2}-\frac{\alpha_n}{C_n\, T_P^n}\left[8-\ln(216)\right]\, ,
\end{align}
where $n=(1,2)$, and we have $A_n=(144\pi^2 Q^2, 432\sqrt{6}\pi^3Q^3)$, $B_n=(24\pi,72\sqrt{6}\pi^2 Q)$, $C_n=(576\sqrt{6}\pi^2 Q^3,10368\pi^3 Q^4)$. The quantities between square brackets in the above equations are positive.
\par
From these expressions, we immediately verify the role of the Planck scale in increasing or reducing the critical temperature depending on the sign of the dimensionless parameter $\alpha_n$, which on the other hand, is proportional to the parameter $\xi$ that is present in the modified dispersion relation (\ref{mdr}). Regardless of the order of the first Planck-scale correction $n$, we see that negative values of $\xi$ are responsible for the phenomenology of investigating an upper threshold in the energy of fundamental particles, which from this approach is reflecting in a raising of the critical temperature. Besides that, $\xi<0$ describes the subluminal propagation of massless particles, i.e., UV photons are propagating with a speed $v<1$. So, we can conclude that the critical temperature, from which phase transition can occur, is raised if we have Planck-scale subluminal propagation. The opposite behavior occurs for superluminal propagation $\xi>0$.
\par
\begin{figure}[h]
    \centering
    \includegraphics[scale=.90]{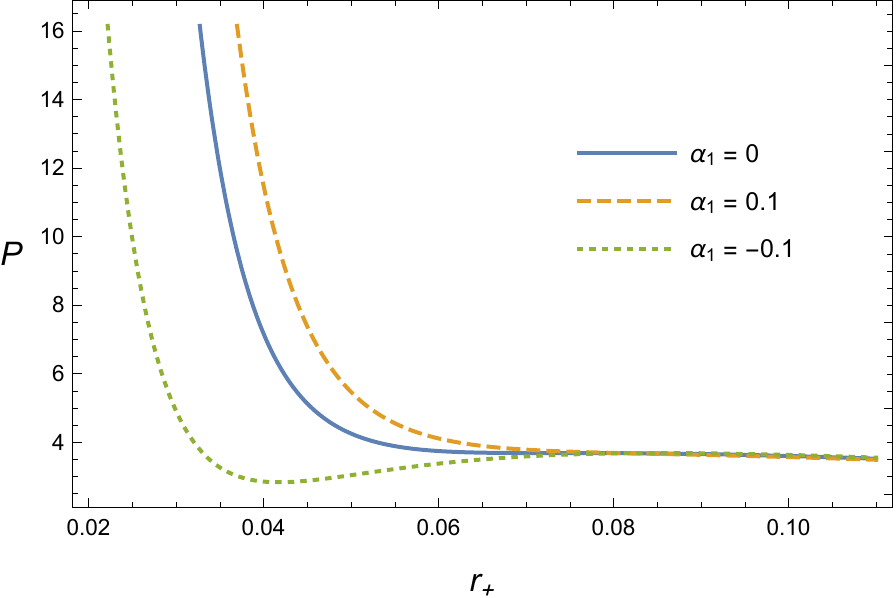}
    \caption{Pressure versus horizon radius for $n=1$ and $Q=0.03$ at constant temperature, $T_c(\alpha_1=0)=\sqrt{6}/18\pi Q$. We assumed natural units in which $T_P=1$. Here it can be seen the induction of criticality by the parameter $\alpha_1$.}
    \label{fig:Pxr1}
\end{figure}
\begin{figure}
    \centering
    \includegraphics[scale=.90]{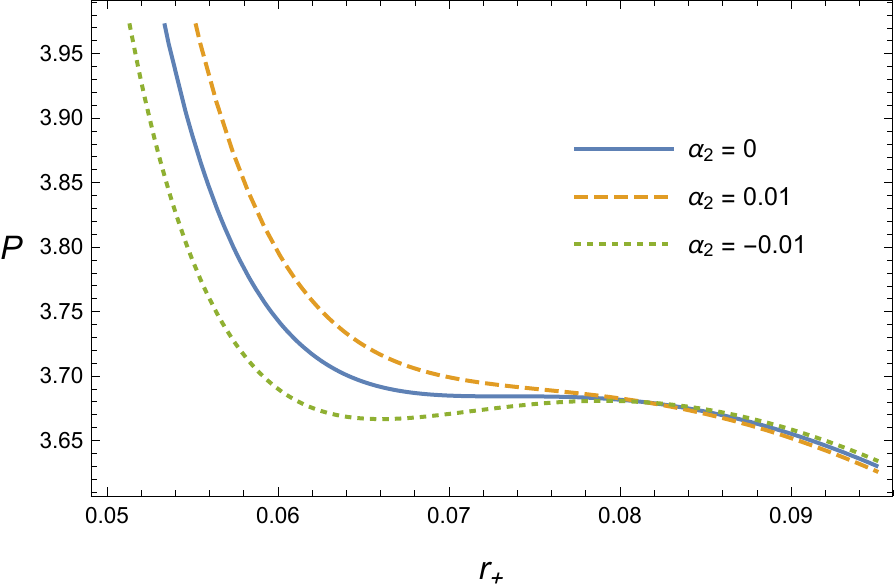}
    \caption{Pressure versus horizon radius for $n=2$ and $Q=0.03$ at constant temperature, $T_c(\alpha_1=0)=\sqrt{6}/18\pi Q$. We assumed natural units in which $T_P=1$. Here it can be seen the induction of criticality by the parameter $\alpha_2$.}
    \label{fig:Pxr2}
\end{figure}

This behavior can be seen in Figs. (\ref{fig:Pxr1}) and (\ref{fig:Pxr2}), where we depicted the pressure as a function of the outer horizon for some values of the quantum gravity parameter $\alpha_n$, assuming the critical temperature $T_c(\alpha_n=0)$.  The blue (solid) curve corresponds to the critical isotherm when $\alpha_n=0$, i.e., it corresponds to a portrayal of the equation of state (\ref{eq:state}) for $T=T_c(\alpha_n=0)$ (where we assumed constant charge). The green (dotted) curve is a depiction of the deformed equation of state parameter (\ref{eq:state}) for $\alpha_n<0$ (subluminal propagation), but assuming the undeformed critical temperature $T_c(\alpha_n=0)$. Since $T_c(\alpha_n=0)<T_c(\alpha_n<0) $, this means that such deformed isotherm lies within the region in the $P-r_+$ diagram that presents phase transitions, which corresponds to temperatures below the critical one for each case. In contrast, the orange (dashed) isotherm corresponds to the case $\alpha_n>0$ (superluminal propagation). Now, it can be seen that $T_c(\alpha_n=0)>T_c(\alpha_n>0)$, which means that this deformed depicted isotherm lies above the critical curve for the superluminal case, thus not manifesting phase transitions.
\par
From the previous considerations, one realizes that the observation or absence of phase transitions in regions of the phase space in which they should be presented can be a manifestation of Planck scale effects, which implies that quantum gravity parameters can be inductors or complicators of phase transitions in Reissner-Nordstr\"{o}m-anti-de Sitter black holes.

\section{Black hole as heat engine}\label{sec:engine}

Thermodynamics was born as a theoretical tool to study the efficiency of heat engines. A necessary component that a thermodynamic system needs to have in order to be analyzed under this perspective is a work term in its conservation of energy equation (first law of thermodynamics). Since the black hole system developed in this paper presents such term due to the identification of the cosmological constant with pressure, it is natural to wonder how this system would behave under a thermodynamic cycle, and what would be its efficiency. This approach, was initially analyzed as a holographic heat engine \cite{Johnson:2014yja} due to the presence of an anti-de Sitter term, that is also a common ingredient of holographic scenarios due to the AdS/CFT conjecture.
\par
A heat engine works between two reservoirs of temperature, a hot and a cold one, and heat flows between these reservoirs. The overall result of this process is work done by the engine at cost at some waste of energy in the form of heat flow to the cold reservoir. This allows one to define the efficiency of the machine as
\begin{equation}\label{efficiency}
\eta= \frac{W}{Q_H}\, ,
\end{equation}
where $W$ is the work done by the system and $Q_H$ is the amount of heat coming from the hot source. The most efficient heat engine is performed following the so called Carnot cycle, constructed by two isothermals and two adiabatics. The most important fact about the Carnot cycle is that the efficiency of the heat engine depends only on the temperature of the reservoirs, and it is given by
\begin{equation}
\eta_C= 1-\frac{T_C}{T_H}\, ,
\end{equation}
where $T_C$ and $T_H$ are the temperatures of the cold and hot reservoirs. As can be seen, the efficiency cannot be equal to unity, since a reservoir cannot have zero temperature.
\par
Since we have a well-defined equation of state involving our thermodynamic variables, given by equation (\ref{T}) and (\ref{tcirc}), the efficiency of the most efficient black hole can be easily obtained. If one wants to calculate the efficiency of an arbitrary cycle, then one needs to use, in general, Eq. (\ref{efficiency}).
\par
\begin{figure}[h]
    \centering
    \includegraphics[scale=.90]{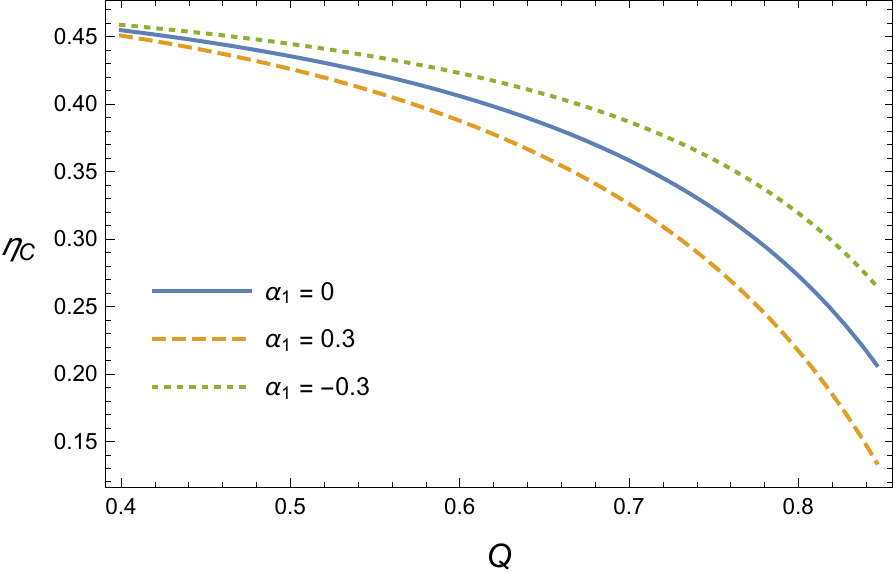}
    \caption{Carnot efficiency versus electric charge for $n=1$. Values of the geometrical quantities are $r_+^C = 1.13$, $P_C = 0.01$ and $r_+^H=0.78$, $P_H = 0.05$. We assumed natural units in which $T_P=1$.}
    \label{fig:carnot1}
\end{figure}
\begin{figure}
    \centering
    \includegraphics[scale=.90]{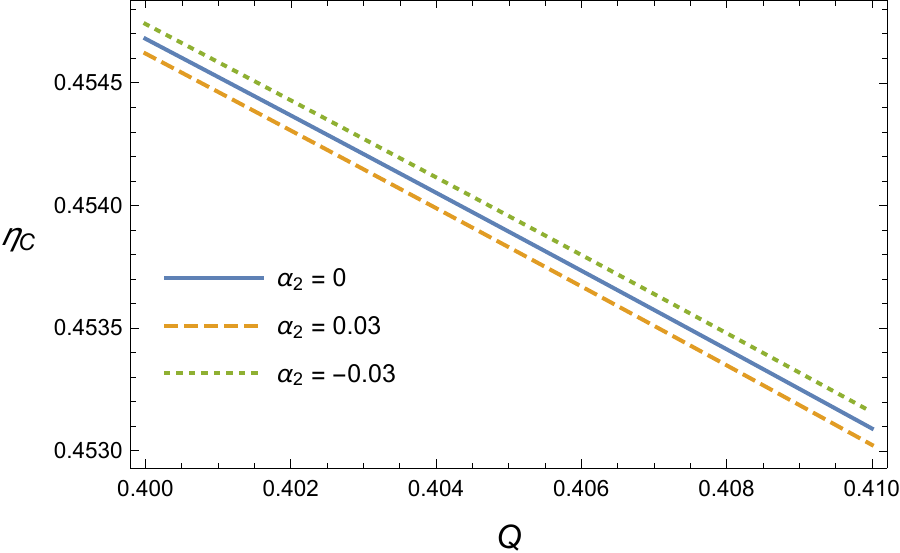}
    \caption{Carnot efficiency versus electric charge for $n=2$. Values of the geometrical quantities are $r_+^C = 1.13$, $P_C = 0.01$ and $r_+^H=0.78$, $P_H = 0.05$. We assumed natural units in which $T_P=1$.}
    \label{fig:carnot2}
\end{figure}

The efficiency of the Carnot cycle for both orders of perturbation ($n=1\, ,\, 2$) is depicted in Figs. (\ref{fig:carnot1}) and (\ref{fig:carnot2}) for some values of the quantum gravity deformation parameter. We are allowed to perform this analysis as long as we do not split too much the three efficiencies in a way that we would need to move to higher order corrections in each case. One sees a universal behavior independently on $n$, in which if UV photons present subluminal propagation ($\alpha_n<0$) the efficiency is raised (green, dotted line), and the opposite behavior occurs for superluminal propagation ($\alpha_n>0$, orange, dashed line). Therefore, Planck scale corrections can improve or worsen the efficiency of the Carnot cycle of the Reissner-Nordstr\"{o}m-anti-de Sitter black hole.

\begin{figure}
\centering
\includegraphics[scale=0.26]{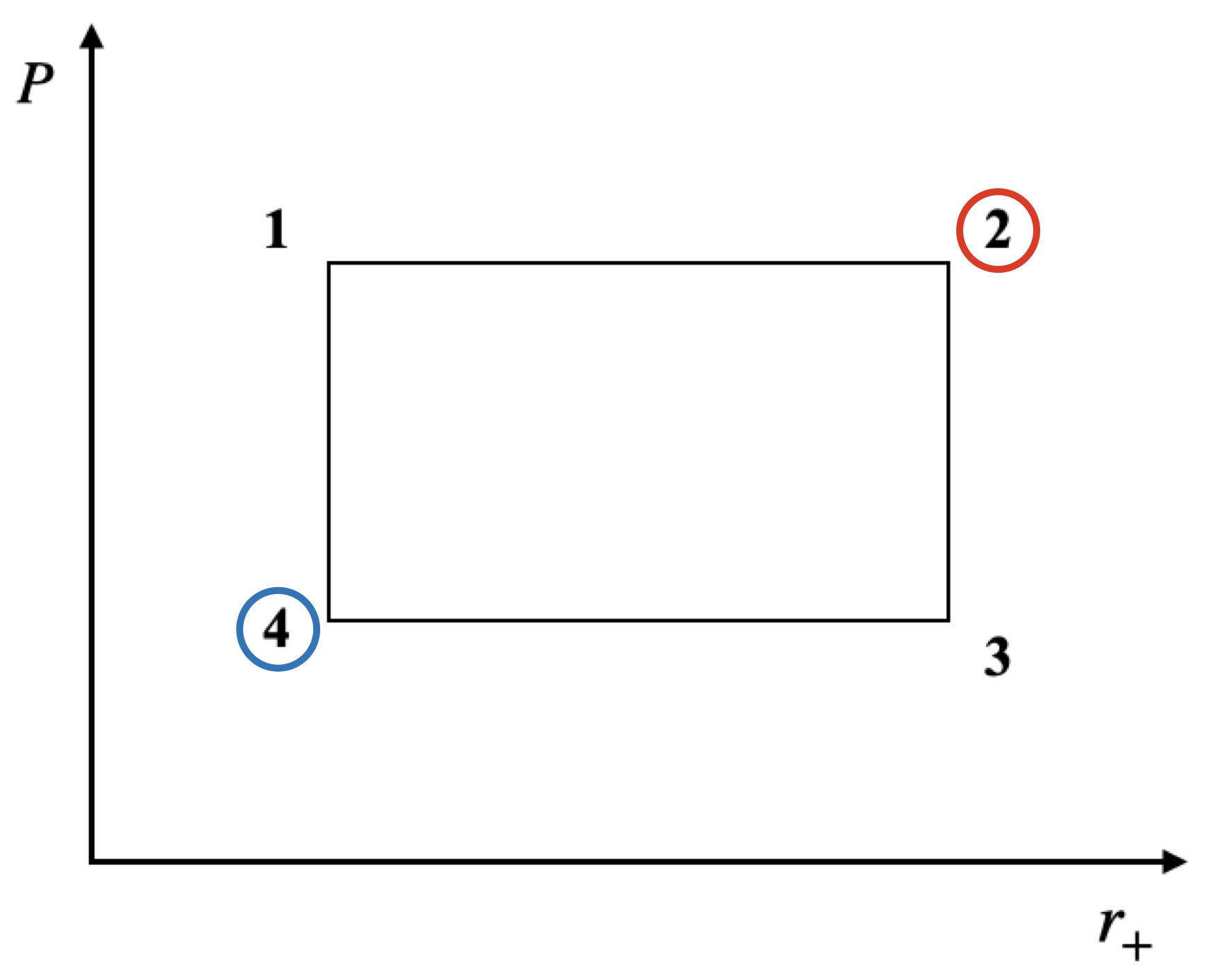}
\caption{A square cycle in the $P-r_+$ diagram. Points $2$ and $4$ are those of highest and lowest temperatures, respectively, between which the black hole as a heat engine works.}\label{square}
\end{figure}

As proposed in the pioneering work \cite{Johnson:2014yja}, we also consider the square circle, depicted in Fig. (\ref{square}). Since the pressure is constant, the work done by the engine is the variation of the enthalpy, which is simply the mass of the black hole. This gives us a straightforward formula for the efficiency as follows:

\begin{equation}
\eta_S= 1-\frac{M_3-M_4}{M_2-M_1}\, .
\end{equation}

\begin{figure}
    \centering
    \includegraphics[scale=.90]{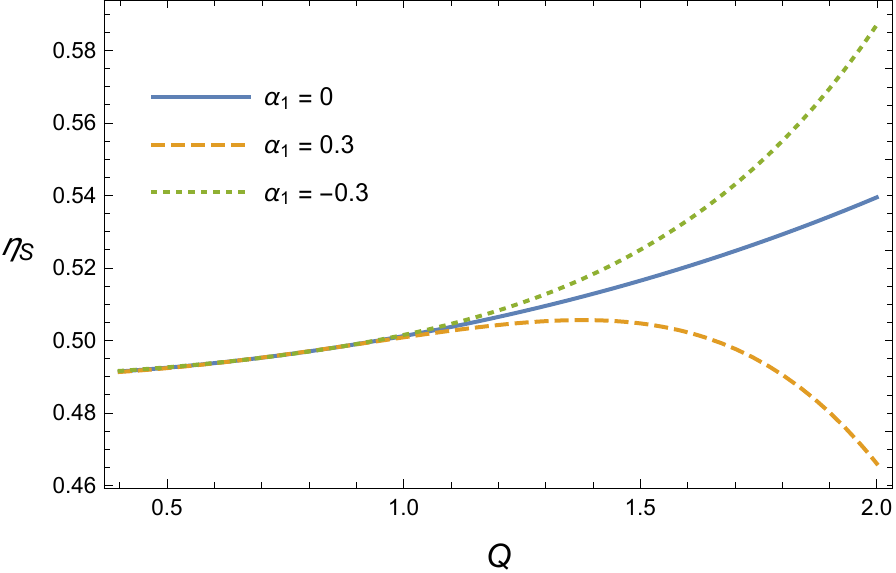}
    \caption{Square cycle efficiency versus electric charge for $n=1$. Values of the geometrical quantities are $r_+^2 = r_+^3 = 1.13$, $r_+^1 = r_+^4 = 0.78$ and $P_1 = P_2 = 2$, $P_3=P_4 = 1$. We assumed natural units in which $T_P=1$.}
    \label{fig:square1}
\end{figure}

For $n=1$, this efficiency is depicted in Fig. (\ref{fig:square1}), and one verifies a different qualitative behavior in comparison to the Carnot cycle. In this case, the subluminal case ($\alpha_1<0$, green, dotted line) is still more efficient, and moves away from the undeformed efficiency as the charge grows. However, the superluminal case ($\alpha_1>0$, orange, dashed line), instead of staying less efficient than the undeformed one (but growing with the electric charge) at a certain point manifests an inflection towards decreasing the efficiency with the growth of the charge. This behavior was not observed in the Carnot cycle, and the point in which not only the superluminal case begins to decrease, but also the subluminal one starts to increase at a more intense way corresponds to $Q\sim 1$ for our choice of points in phase space.
\par
\begin{figure}
    \centering
    \includegraphics[scale=.90]{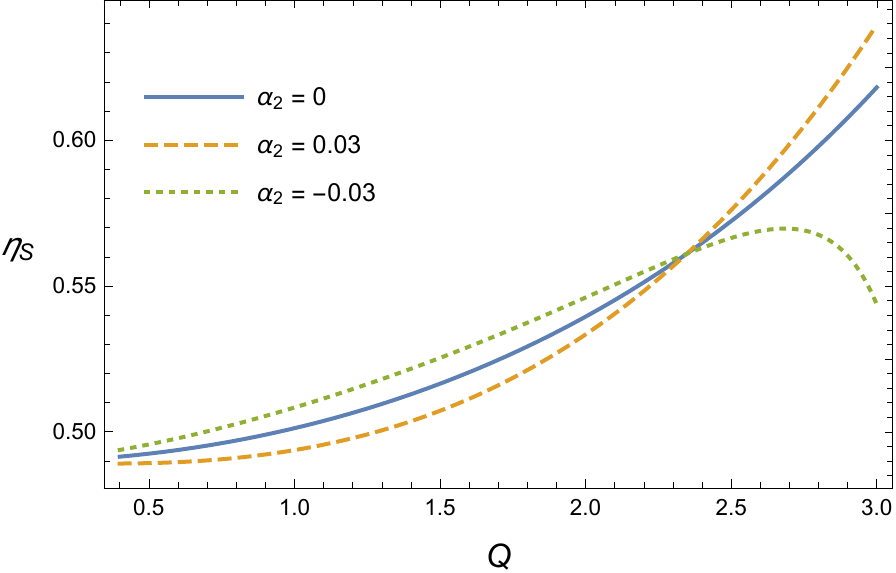}
    \caption{Square cycle efficiency versus electric charge for $n=2$. Values of the geometrical quantities are $r_+^2 = r_+^3 = 1.13$, $r_+^1 = r_+^4 = 0.78$ and $P_1 = P_2 = 2$, $P_3=P_4 = 1$. We assumed natural units in which $T_P=1$.}
    \label{fig:square2}
\end{figure}

The order of the Planck scale perturbation plays a significant role in the square cycle, as can be verified in Fig. (\ref{fig:square2}). In this case, for low charges, we have the same ordering of the previous cases, i.e., with the less efficient engine being the superluminal case ($\alpha_2>0$, orange, dashed line), passing through the undeformed case ($\alpha_2=0$, blue, solid line), until the subluminal case ($\alpha_2<0$, green, dotted line) provides the most efficient heat engine. However, we show here that this is not an universal behavior, since, for the geometric values assigned in this cycle (which were the same considered in Fig. (\ref{fig:square1})) when the black hole's electric charge reaches the value $Q\sim 2.5$, we verify an inversion of the behavior, in which the subluminal case starts to behave in a way similar to the superluminal one. The opposite behavior occurs for the superluminal case, in which $\alpha_2>0$ describes the most efficient way of performing work from the black hole system, and $\alpha_2<0$ becomes the less efficient one.

\section{Final remarks}\label{sec:conc}
Endowed with the Kiselev proposal for describing a black hole surrounded by matter fluids, which would depend on their equation of state parameters, we propose an effective generalization of the Reissner-Nordstr\"{o}m-anti-de Sitter black hole spacetime depending on Planck-scale corrections. This approach allows us to analyze the black hole extended phase space thermodynamics and its quantum gravity effects.
\par
The introduction of a negative cosmological constant as a pressure term allowed us to derive some non-trivial results, like the presence of dissipation and instabilities in the linear regime (a property that is absent in the quadratic one), and logarithmic corrections for some the thermodynamic quantities, like the black hole's electric potential, entropy, thermodynamic volume, temperature and mass.
\par
The extended phase space thermodynamics of the Reissner-Nordstr\"{o}m-anti-de Sitter spacetime is sometimes referred as black hole chemistry, due to the resemblance of its equations of state with a Van der Waals gas, which are able to manifest phase transitions. In fact, we analyzed the effect of the quantum gravity parameters in the critical behavior of this black hole system assuming linear and quadratic Planck scale contributions. Among our findings, we verified that if UV photons have subluminal (superluminal) propagation, the critical temperature below which phase transitions can occur is raised (lowered). Besides that, critical radius and pressure were calculated.
\par
We also analyzed the black hole as a heat engine and found that the efficiency of the Carnot cycle presents properties that are common to the linear and quadratic corrections by raising (lowering) the efficiency of this engine if UV photons present subluminal (superluminal) propagation. However, this is not a universal behavior for all cycles, as we demonstrated using the square cycle in a $P-r_+$ diagram, in which the linear case presents a more drastic splitting in the behavior of the efficiency for a given critical value of the electric charge (however still obeying the same hierarchy of efficiencies of the Carnot cycle), and the quadratic case presents an inversion in the behavior of the efficiency. Therefore, at a given value of the electric charge, the hierarchy of efficiencies reverses and the superluminal case becomes dominant over the undeformed and subluminal ones, respectively.
\par
The analyses done in the paper show that Planck scale corrections on the thermodynamics of the radiation that works as a source of the Reissner-Nordstr\"{o}m-anti-de Sitter black hole can present unexpected properties: like pressure-driven dissipation for linear corrections, logarithmic contributions in other thermodynamic quantities, not only the entropy, deformations of the phase transition regions, and alternating efficiency hierarchies involving subluminal and superluminal propagation of UV photons depending on the nature of the Planck-scale corrections (if linear or quadratic) and on the considered cycles.

\section*{Acknowledgements}
I. P. L. would like to acknowledge the contribution of the COST Action CA18108. I. P. L. was partially supported by the National Council for Scientific and Technological Development - CNPq grant 306414/2020-1. J. P. M. G. was partially supported by the National Council for Scientific and Technological Development - CNPq grant 151701/2020-2. V. B. B. was partially supported by the National Council for Scientific and Technological Development - CNPq grant 307211/2020-7. 

\end{document}